\documentclass[twocolumn,showpacs,aps,prl,superscriptaddress]{revtex4}
\usepackage{multirow}
\usepackage{graphicx}
\usepackage{dcolumn}
\usepackage{amsmath}
\usepackage{epsfig}
\usepackage{hyperref}
\input{ogbabarsym}

\newcommand{\BABARPubYear}    {12}
\newcommand{\BABARPubNumber}  {006}

\newcommand{\SLACPubNumber} {15087}

\def\figurebox#1#2#3{%
    \def\arg{#3}%
    \ifx\arg\empty
    {\hfill\vbox{\hsize#2\hrule\hbox to #2{\vrule\hfill\vbox to #1{\hsize#2\vfill}\vrule}\hrule}\hfill}%
    \else
    {\hfill\epsfbox{#3}\hfill}%
    \fi}

\begin{document}

\preprint{\babar-PUB-\BABARPubYear/\BABARPubNumber}
\preprint{SLAC-PUB-\SLACPubNumber} 

\begin{flushleft}
\babar-PUB-\BABARPubYear/\BABARPubNumber\\
SLAC-PUB-\SLACPubNumber\\
\end{flushleft}

\title{\vspace{2em}{\large \bf\boldmath Study of the baryonic \B decay $\mydec$\unboldmath}}


%
\author{J.~P.~Lees}
\author{V.~Poireau}
\author{V.~Tisserand}
\affiliation{Laboratoire d'Annecy-le-Vieux de Physique des Particules (LAPP), Universit\'e de Savoie, CNRS/IN2P3,  F-74941 Annecy-Le-Vieux, France}
\author{J.~Garra~Tico}
\author{E.~Grauges}
\affiliation{Universitat de Barcelona, Facultat de Fisica, Departament ECM, E-08028 Barcelona, Spain }
\author{A.~Palano$^{ab}$ }
\affiliation{INFN Sezione di Bari$^{a}$; Dipartimento di Fisica, Universit\`a di Bari$^{b}$, I-70126 Bari, Italy }
\author{G.~Eigen}
\author{B.~Stugu}
\affiliation{University of Bergen, Institute of Physics, N-5007 Bergen, Norway }
\author{D.~N.~Brown}
\author{L.~T.~Kerth}
\author{Yu.~G.~Kolomensky}
\author{G.~Lynch}
\affiliation{Lawrence Berkeley National Laboratory and University of California, Berkeley, California 94720, USA }
% ...
\collaboration{The \babar\ Collaboration}
\noaffiliation


\begin{abstract}
We report the measurement of the baryonic \B decay $\mydec$. Using a data sample of 
$467 \times 10^6$ \BB pairs collected with the \babar\ detector at the \pep2 \ storage 
ring at SLAC, the measured branching fraction is 
$(2.98 \pm 0.16_{\stat} \pm 0.15_{\syst} \pm 0.77_{(\rm \Lambda_c)}) \times 10^{-4}$,
where the last error is due to the uncertainty in $\BR(\Lcp\to\proton\Km\pip)$.
The data suggest the existence of resonant subchannels $\Bub\to\LcpsI\antiproton\pim$
and, possibly, $\mydecresA$. We see unexplained structures in $m(\Scpp\pim\pim)$ at $3.25\gevcc$, 
$3.8\gevcc$, and $4.2\gevcc$.
\end{abstract}


\maketitle


\begin{center}
 \textbf{I. Introduction}
\end{center}

The large mass of the \B meson allows a wide spectrum of baryonic decays, which have, 
in total, a branching fraction of $(6.8 \pm 0.6)\,\%$ \cite{ref:PDG}.
This makes \B decays a good place to study the mechanisms of baryon production.
One approach to investigate the baryonization process in \B decays is to measure 
and compare their exclusive branching fractions and study the dynamic structure of 
the decay, i.e., the influence of resonant subchannels.


In this paper, we present a study of the decay $\mydec$ \cite{footnote}. This decay is a resonant 
subchannel of the five body final state \mydecF, which has the largest hitherto known branching 
fraction among all baryonic \B decays and hence is a good starting point for further investigations.
The analyzed decay can be compared with \mydecS and \deca, which have similar quark content
and phase space.

Large differences between the branching fractions of \mydec, \mydecS, and \deca could indicate 
a considerable impact of intermediate states on baryonic \B decays. For example the decay \mydecS
allows a number of resonant three-body decays (including $\Nbar$, $\Deltabar^{0}$, and $\rho^{0}$
resonances) that cannot occur in \mydec.
The importance of resonant subchannels can be quantified, e.g., by the ratio of
$\left[\BR(\mydec)+\BR(\mydecS)\right]/\BR(\mydecF)$.

The CLEO Collaboration measured $\BR(\mydecF)=(22.5 \pm 3.5 \pm 5.8)\times10^{-4}$ and 
$\BR(\mydecS)=(4.4 \pm 1.7 \pm 1.1)\times10^{-4}$ \cite{ref:CLEO}. 
The decay \deca was measured by the CLEO \cite{ref:CLEO} and the Belle \cite{ref:Belle2} Collaborations.
The Particle Data Group has calculated an average of $\BR(\deca)=(2.2 \pm 0.7 \pm 0.6)\times10^{-4}$ 
\cite{ref:PDG}. For all these branching fractions the first uncertainty is the combined statistical and 
systematic error and the second one is due to the uncertainty in $\lcpdecbr=(5.0\pm1.3)\,\%$ \cite{ref:PDG}.


\begin{center}
 \textbf{II. The \babar\ experiment}
\end{center}

This analysis is based on a dataset of about $426 \invfb$, corresponding to $467 \times 10^6$ \BB pairs.
The sample was collected with the \babar\ detector at the \pep2 \ asymmetric-energy 
\epem storage ring, which was operated at a center-of-mass energy equal to the \FourS mass. 
For generation of Monte Carlo (MC) simulated data we use \evtgen \cite{ref:EvtGen} for event 
generation and \geant4 \cite{ref:geant} for detector simulation.


The \babar\ detector is described in detail elsewhere \cite{ref:NIM}.
The selection of proton, kaon, and pion candidates is based on measurements of the energy loss 
in the silicon vertex tracker and the drift chamber, and of the Cherenkov radiation in the detector of 
internally reflected Cherenkov light \cite{Aubert:2002rg}.
The average efficiency for pion identification is approximately $95\,\%$, with a typical misidentification
rate of $10\,\%$ due to other charged particles such as muons and kaons, depending on the momentum and 
the polar angle of the particle. The efficiency for kaon identification is about $95\,\%$ with a 
misidentification rate less than $5\,\%$ due to protons and pions. 
The efficiency for proton and antiproton identification is about $90\,\%$ with a misidentification rate 
about $2\,\%$ due to kaons.

\begin{center}
 \textbf{III. Decay reconstruction}
\end{center}


The decay $\mydec$ is reconstructed in the subchannel $\Scpp\to\Lcp\pip$, $\Lcp\to\proton\Km\pip$.
For the reconstruction of the \B candidate the entire decay tree is fitted simultaneously.
A vertex fit is performed for \Bub, \Scpp, and \Lcp, and the $\chi^2$ fit probability 
is required to exceed $0.1\,\%$.


To suppress background, the invariant mass of the $p\,\Km\pip$ combination
is required to satisfy $2275\mevcc<m_{\proton\Km\pip}<2296\mevcc$, i.e., compatible with 
coming from the decay \Lcp\to\proton\Km\pip. This selection corresponds to $2.8$ times the 
observed width of reconstructed \Lcp\ candidates which are centered at $m_{\proton\Km\pip}=2285.4\mevcc$.
The separation of signal from background in the \B-candidate sample is obtained using two 
kinematic variables, $\DeltaE = E_{\rm \B}^* - \sqrt{s}/2$ and 
$\mes = \sqrt{(s/2+\mathbf{p}_{\rm i} \cdot \mathbf{p}_{\rm \B})^2/E_{\rm i}^2-\left|\mathbf{p}_{\rm \B}\right|^2}$, 
where $\sqrt{s}$ is the center-of-mass (CM) energy of the \epem pair and $E_{\rm \B}^*$ the energy of the 
\B candidate in the CM system. $(E_{\rm i}, \mathbf{p}_{\rm i})$ is the four-momentum vector of the \epem CM 
system and $\mathbf{p}_{\rm \B}$ the \B-candidate momentum vector, both measured in the laboratory frame.
For correctly reconstructed \B decays, \mes is centered at the \B meson mass and \DeltaE is centered 
at zero. Throughout this analysis, \B candidates are required to have \mes within $8\mev$ $(3.4\sigma)$ 
of the measured \B mass of $\mes=5279.1\mev$.

\begin{figure}[ht!]
	\centering\includegraphics[width=.5\textwidth]{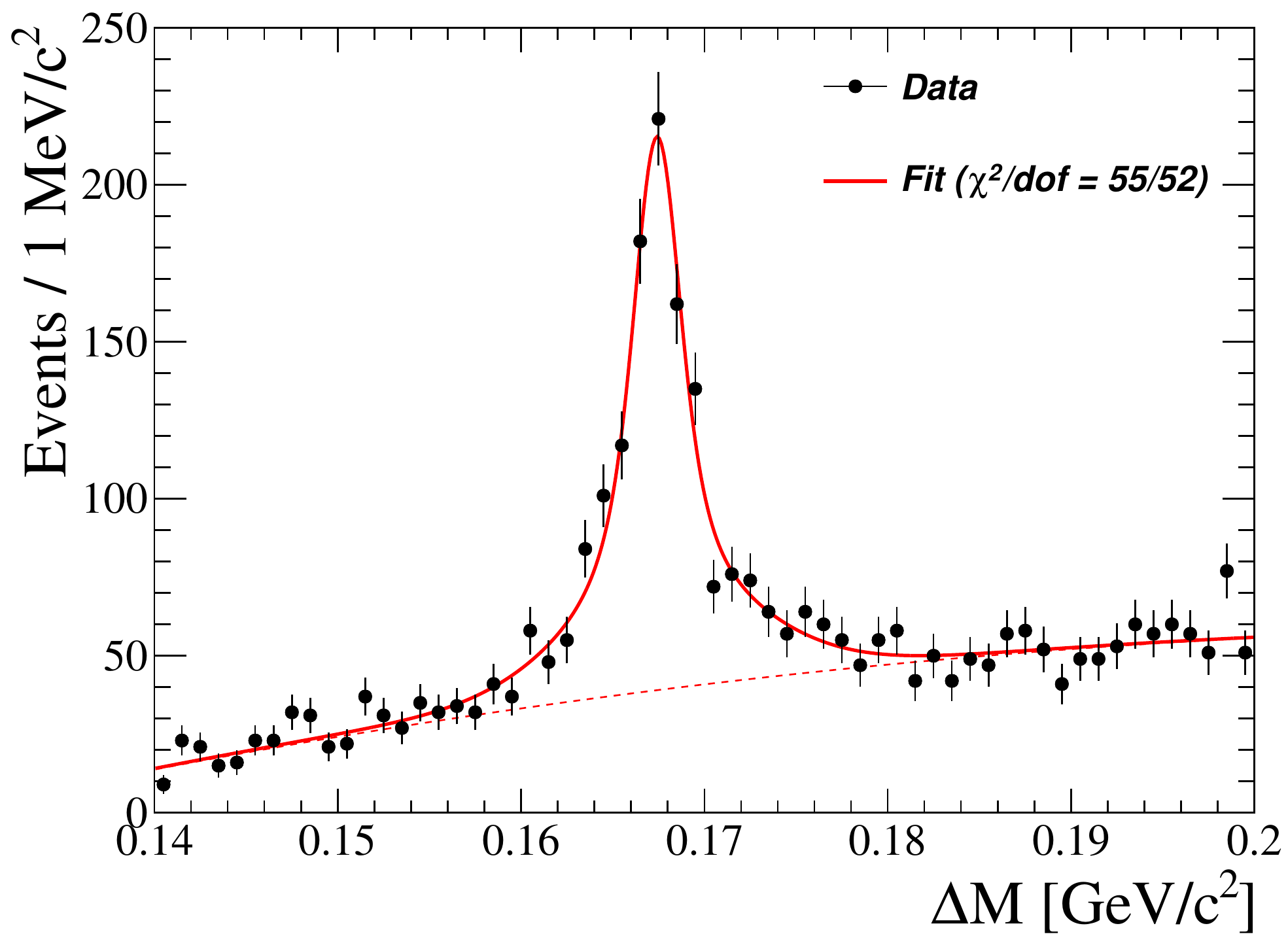}
	\caption{Fitted \mdiff distribution for \B candidates in data. All candidates are 
	required to satisfy the selection criteria on \mes, $m_{\proton\Km\pip}$, and \DeltaE.}
	\label{fig:1}
\end{figure}

Figure~\ref{fig:1} shows the distribution of $\mdiff\equiv m(\Lcp\pip)-m(\Lcp)$ in data 
for candidates that satisfy the criteria described above for \mes and $m_{\proton\Km\pip}$ 
and for which \DeltaE is between $-60\mev$ and $+40\mev$. We perform a binned minimum \chisq 
fit using a second-order polynomial for the description of the background and the sum of a 
Voigt distribution (the convolution of a Breit-Wigner function with a Gaussian function) and 
a Gaussian to parametrize the \Scpp\ signal. A detailed explanation of the fit function is 
given in Sec. IV. The fitted \Scpp\ signal yield is $N=1020\pm95$.

\begin{figure}[ht!]
	\centering\includegraphics[width=.5\textwidth]{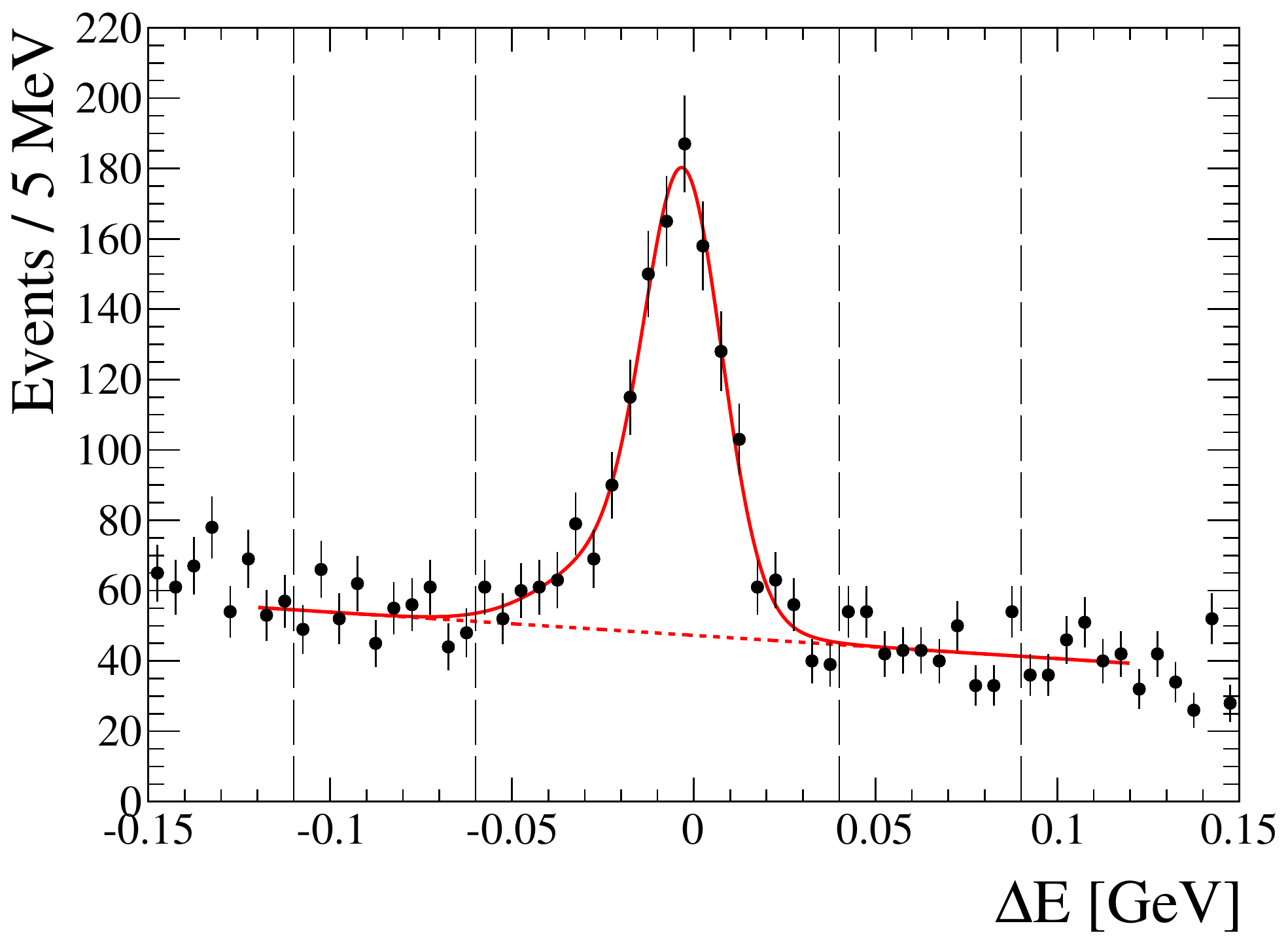}
	\caption{Fitted \DeltaE distribution in data with selection criteria applied to
	$m_{\proton\Km\pip}$, \mes, and \mdiff. The goodness of the fit is $\chisq/{\rm dof}=36/40$.
	The \DeltaE signal region is between $-60\mev$ and $40\mev$, and enclosed by the two sideband 
	regions that each have a width of $50\mev$.}
	\label{fig:2}
\end{figure}

Figure~\ref{fig:2} shows the \DeltaE distribution in data for candidates that
satisfy the criteria described above for \mes and $m_{\proton\Km\pip}$ and for 
which \mdiff is between $0.157\gevcc$ and $0.178\gevcc$. The latter is a selection 
of \Scpp\ candidates with an efficiency of $92\%$ in signal MC.

In the binned minimum \chisq fit we use the sum of two Gaussian functions for the signal and a 
linear function for the background. 
The second Gaussian accommodates \B decays with missing energy due to final state radiation.
Each Gaussian has a mean parameter ($\mu$) and a standard deviation ($\sigma$). 
The joint normalization is described by $N_{\rm sig}$ and the fraction of the first Gaussian 
is $f_{1}$. We parametrize the background shape of the \DeltaE distribution as a first-order 
polynomial which provides a good description of the \DeltaE distribution for candidates in 
the \mes sideband in the range $5.20\gevcc<\mes<5.26\gevcc$.
All parameters are permitted to vary during fitting. Table \ref{tab:par} presents the resulting 
signal parameters. The signal yield is $840\pm55$ events.

\begin{table}[ht!]
\caption{The parameters for the double-Gaussian function describing the signal contribution 
in the fit to the \DeltaE distribution shown in Fig. \ref{fig:2}.
$f_{1}$ is the fraction of the signal in the narrower Gaussian.}
\vspace{0.5em}
\begin{tabular}{lr}
\toprule 
\rule{0mm}{4mm} Parameter \hspace{0.45\linewidth} & Fit result \\
\hline
\rule{0mm}{4mm} $N_{\rm sig}$ & $840\pm55$ \\ 
\rule{0mm}{4mm} $f_{1}$ & $(70\pm23)\%$ \\ 
\rule{0mm}{4mm} $\mu_{1}$ & $(-2.7\pm1.2)\mev$ \\ 
\rule{0mm}{4mm} $\sigma_{1}$ & $(10\pm1.6)\mev$ \\ 
\rule{0mm}{4mm} $\mu_{2}$ & $(-16\pm14)\mev$ \\ 
\rule{0mm}{4mm} $\sigma_{2}$ & $(20\pm5.6)\mev$ \\ 
\toprule 
\end{tabular}
\label{tab:par}
\end{table}


\begin{center}
 \textbf{IV. Signal extraction}
\end{center}

There are two sources of background that contribute to the signal in \DeltaE
and \mdiff. The first one is \B decays that have the same final state, in particular
\mydecF, and the other one is \B decays that have a \Scpp\ among its decay products,
e.g., \Bzb\to\Scpp\antiproton\pim\piz.
To reject this background we make a binwise fit using \mdiff as a discriminating variable 
to create a background-subtracted \DeltaE distribution from which we extract the true 
signal yield in order to determine \mybr. The binwise fitting procedure is described 
in the following paragraph.

After applying the selection in $m_{\proton\Km\pip}$ and $\mes$ (no selection in \mdiff), we divide 
the $\DeltaE$ range $(-105,105)\mev$ into 14 equal slices and fit the $\mdiff$ 
distribution in each slice separately in the range $0.14\gevcc<\mdiff<0.2\gevcc$.
In the fits the \Scpp\ signal is represented by the sum of a Voigt function 
and a Gaussian function. The Voigt distribution has four parameters ($N_{\rm sig}, \mu, \Gamma, \sigma$)
and models the signal peak region, where $\mu$ is the mean of the Voigt and 
represents the \Scpp\ mass, which is fixed to the value obtained from an 
inclusive analysis of $\Scpp\to\Lcp\pip$ candidates in the data. The parameter $\Gamma$
is the intrinsic width of the $\Scpp$ and is fixed to the 2010 Review of Particle Properties (RPP)
value \cite{ref:PDG}, and $\sigma$ describes the detector resolution in \mdiff for the \Scpp\ 
determined, independently for each $\Delta E$ slice, from the signal MC. The remaining 
parameter $N^{i}_{\rm sig}$ is the fitted \Scpp\ signal yield in each of the \DeltaE bins.

There is a correlation between \mdiff and \DeltaE that is very prominent due to the inaccurate momentum 
measurement of the slow \pip from the $\Scpp$ decay. As a result the \Scpp\ signal has tails in the 
\mdiff distribution that are modeled by the Gaussian function whose parameters are determined,
independently for each $\Delta E$ slice, from the signal MC.
The background is represented by a second-order polynomial.
This shape was determined from the sidebands $|\DeltaE|\in(50,300)\mev$ and, 
compared to the other polynomials, gives the best \chisq fit probability.
The fits in \mdiff determine the background level and the number of $\Scpp$ baryons.

\begin{figure}[ht!]
	\centering\includegraphics[width=.5\textwidth]{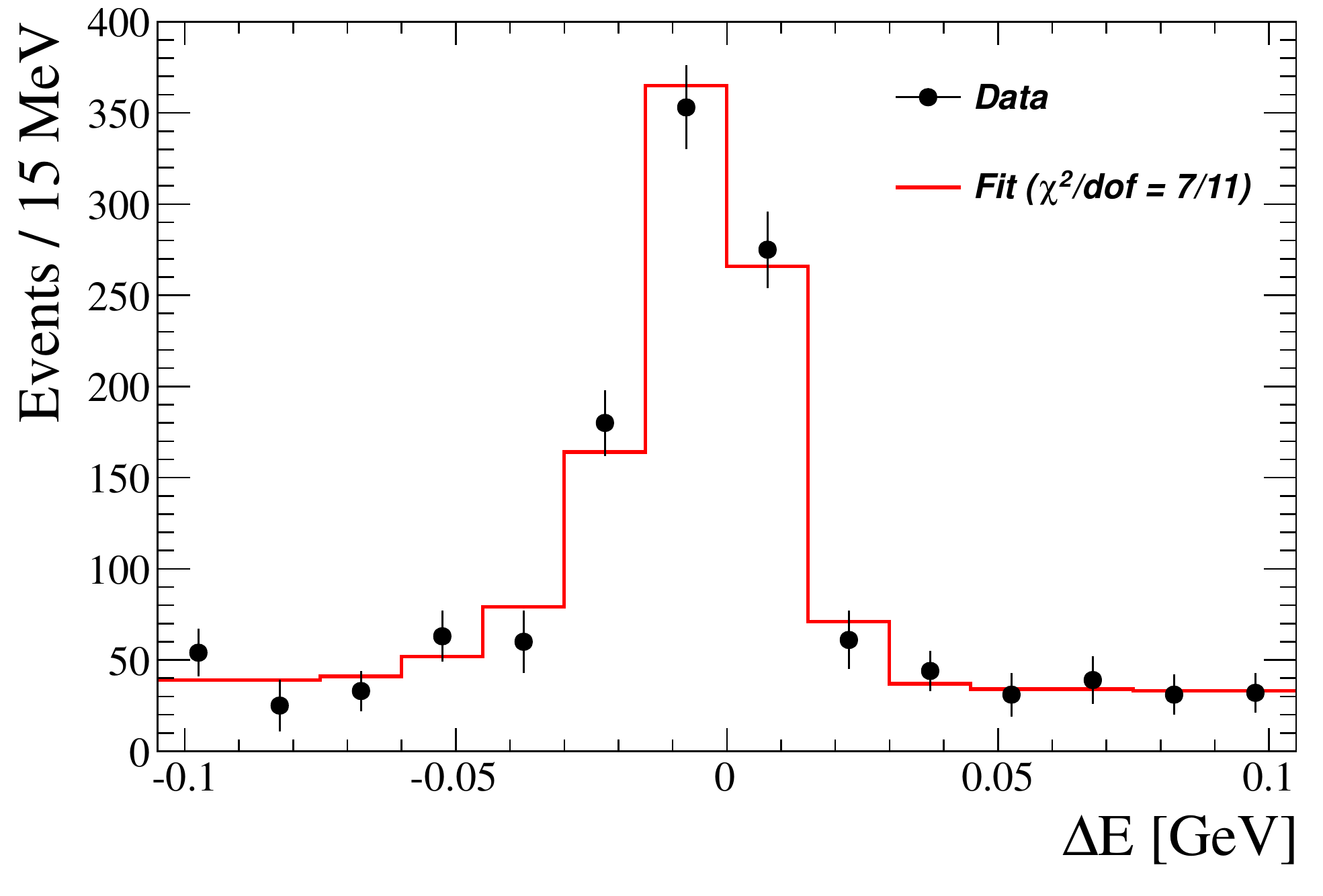}
	\caption{\DeltaE distribution for \mydec candidates in data. The points with error bars
	represent the number of \Scpp\ candidates $N_{\rm sig}^{i}$ from a fit to $\mdiff\equiv\diffmass$. 
	All signal parameters for the \DeltaE distribution, except $N_{\rm sig}$, are fixed to
	those shown in Table \ref{tab:par}.}
	\label{fig:3}
\end{figure}

Figure~\ref{fig:3} shows the \Scpp\ signal yield as a function of \DeltaE. We fit this distribution 
with the same functions described in Sec. III and fix the signal parameters, except for $N_{\rm sig}$, 
to those determined there. The true signal yield is $N_{\rm sig} = 787\pm 43$ events.

\begin{center}
 \textbf{V. Efficiency}
\end{center}

The efficiency is calculated from the simulated events. These events were generated uniformly in 
four-body phase space (PS), but the actual decay distribution is, a priori, unknown. Therefore, 
when calculating the efficiency, we weight the MC events so that we reproduce the distributions of
the two-body invariant mass distributions for the decay products of the \B candidates in data.
The resulting efficiency is checked by repeating the procedure using the three-body masses and 
then again using the angles between the \B daughters in the \B rest frame. The different procedures
give an average efficiency of $(11.3\pm0.2_{\syst})\,\%$, which is used to determine the branching 
fraction. Out of the efficiencies from the different procedures, we use the maximum deviation from the 
average efficiency as systematic uncertainty. The statistical uncertainty, due to the use of the data, 
is negligible compared to the statistical uncertainty in the event yield. The efficiency calculated 
using unweighted events is $11.0\,\%$.


\begin{center}
 \textbf{VI. Systematic uncertainties}
\end{center}

We estimate the uncertainty on the signal extraction in three different ways:
(1) the fit to \DeltaE in Fig.~\ref{fig:3} is repeated separately for each shape 
parameter in Table \ref{tab:par}, while permitting this parameter to float. The 
absolute deviations ($\delta N$) in the event yield to our true signal yield 
$N_{\rm sig} = 787$ add up to $23$ (see Table \ref{tab:sys}).
(2) We use a second-order polynomial for the background while letting all other 
parameters fixed ($\delta N=5$), and (3) we fit only the background with a first-order 
polynomial and subtract its integral from the histogram content in the range 
$-60\mev<\DeltaE<45\mev$ in order to obtain an alternative signal yield ($\delta N=3$). 
The absolute values of the deviations in the event yields from all of these variations
add up to $31$. The resulting relative uncertainty on the signal yield is $4.0\,\%$. 
Other systematic errors come from track reconstruction efficiency $(2.4\,\%)$ \cite{babar:trackfinding}, 
efficiency $(1.8\,\%)$, and the number of produced \BB pairs in the data sample $(1.1\,\%)$. 
The total relative uncertainty on the branching fraction is $5.1\,\%$.

\begin{table}[ht!]
\caption{The results of the fits to \DeltaE in Fig.~\ref{fig:3} while the given parameter is allowed to float.
$\delta N$ is the absolute deviation to our true signal yield $N_{\rm sig} = 787$.}
\vspace{0.5em}
\begin{tabular}{lcr}
\toprule 
\rule{0mm}{4mm} Floating parameter & \hspace{0.18\linewidth} Fit result \hspace{0.18\linewidth} & $\delta N$ \\
\hline
\rule{0mm}{4mm} $f_{1}$ & $(70\pm7.7)\%$ & \quad $2$ \\ 
\rule{0mm}{4mm} $\mu_{1}$ & $(-2.8\pm1.0)\mev$ & \quad $0$ \\ 
\rule{0mm}{4mm} $\sigma_{1}$ & $(11\pm0.9)\mev$ & \quad $8$ \\ 
\rule{0mm}{4mm} $\mu_{2}$ & $(-15\pm5.2)\mev$ & \quad $2$ \\ 
\rule{0mm}{4mm} $\sigma_{2}$ & $(18\pm4.6)\mev$ & \quad $11$ \\ 
\toprule 
\end{tabular}
\label{tab:sys}
\end{table}


\begin{center}
 \textbf{VII. Branching fraction results}
\end{center}

Using the results from the signal extraction, efficiency determination, and estimation of 
systematic errors we find

\vspace{-1.0em}

$$\BR(\mydec)\cdot\BR(\Lcp\to\proton\Km\pip) = \frac{N_{\rm sig}}{\varepsilon\cdot N_{\rm \BB}}$$

\vspace{-1.0em}

\begin{equation}
= (1.49 \pm 0.08_{\stat} \pm 0.08_{\syst})\times10^{-5}\, ,\, \text{and}
\end{equation}

\vspace{-1.0em}

$$\BR(\mydec) = \frac{N_{\rm sig}}{\varepsilon\cdot N_{\rm \BB}\cdot\BR(\Lcp\to\proton\Km\pip)}$$

\begin{equation}
\label{eq:2}
=\left(2.98\pm0.16_{\stat}\pm0.15_{\syst}\pm0.77_{(\rm \Lambda_{c})}\right)\times10^{-4}.
\end{equation}

\noindent
In Eq. \ref{eq:2} the last error is due to the uncertainty in $\BR(\Lcp\to\proton\Km\pip)$.


\begin{center}
 \textbf{VIII. Fraction of PS distributed decays}
\end{center}

\begin{figure}[ht!]
	\centering\includegraphics[width=.5\textwidth]{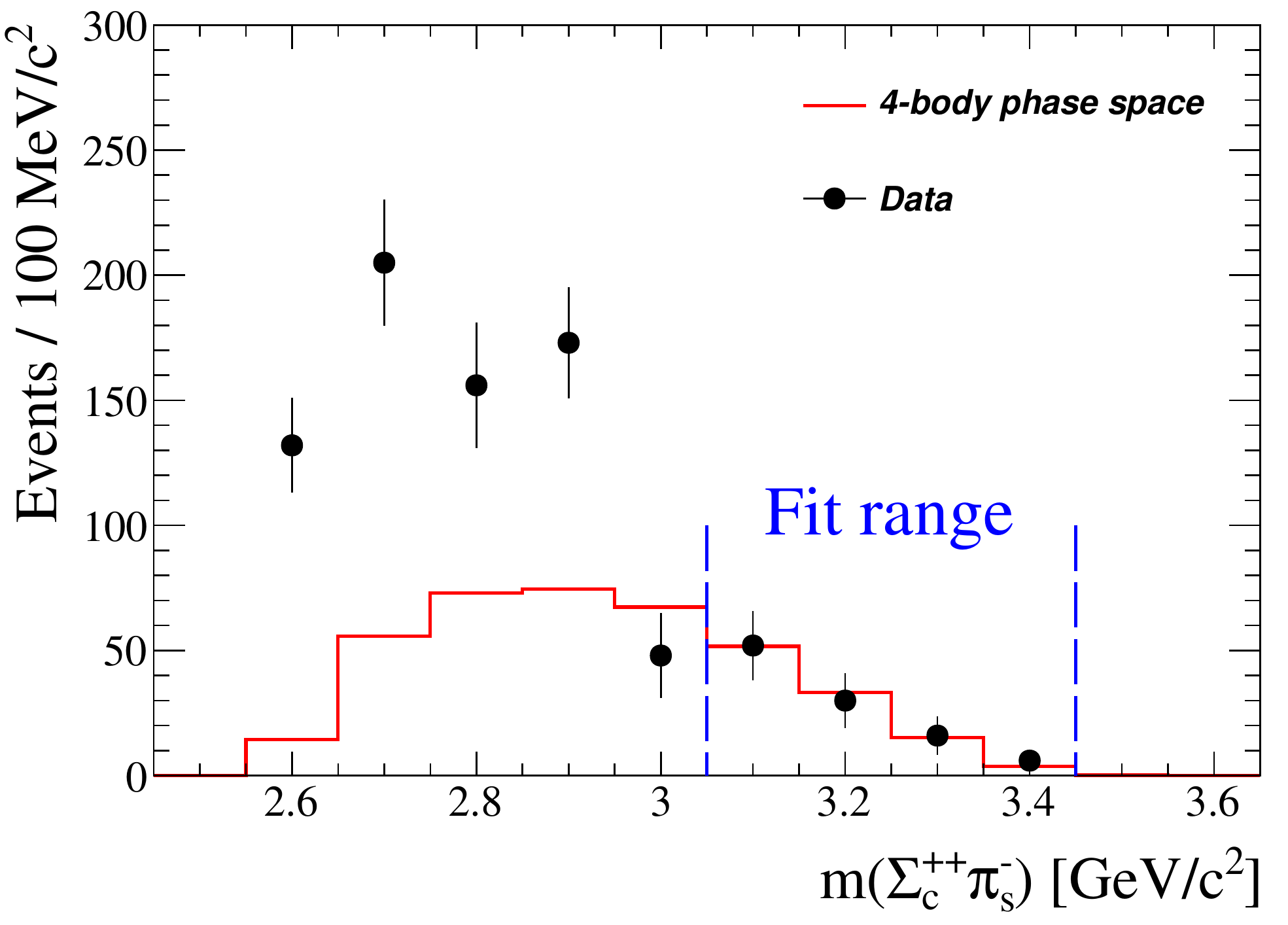}
	\caption{The $m(\Scpp\pislow)$ distribution in data (points with error bars)
	and for simulated four-body phase space decays (histogram).
	The distribution in data results from a sideband subtraction in \DeltaE according to
	the definition in Fig.~\ref{fig:2}.}
	\label{fig:4}
\end{figure}

To compare the two-body and three-body invariant masses of the \B decay products
in data with PS, we determine an effective PS fraction of the total branching ratio. 
To do this, we assume that the resonant substructures are due to the intermediate states 
$\Lcps\to\Scpp\pim$ and $\Deltabar^{--}\to\antiproton\pim$, and the remainder is distributed
according to four-body PS.
We investigate all two-dimensional planes that are spanned by the two-body invariant
masses of the \B decay products, e.g. $m(\Scpp\pislow)$ against $m(\antiproton\pifast)$,
to look for a range that is free from \Lcps and $\Deltabar^{--}$ resonances and hence can
be described by a four-body PS distribution. The symbol \pislow refers to the $\pim$ that has
the lower momentum in the $\epem$ CM system. The other $\pim$ is denoted as \pifast.
We see no indication of $\Deltabar^{--}$ and $\Lcps$ resonances for \B candidates in the 
range $3.050\gevcc<m(\Scpp\pislow)<3.450\gevcc$, where the normalization of the PS 
distribution is determined by fitting the sideband-subtracted data (Fig. \ref{fig:4}).

From the ratio of the efficiency-corrected integrals of the distributions in Fig.\ref{fig:4}, 
we calculate an effective PS fraction:

\vspace{-0.5em}

\begin{equation}
\label{eq:3}
\dfrac{\mybr_{\rm PS}}{\mybr}=\dfrac{389}{11.0\,\%}\cdot\dfrac{11.3\,\%}{816}=49\,\%\,.
\end{equation}

\begin{figure}[htp]
	\centering\includegraphics[width=.5\textwidth]{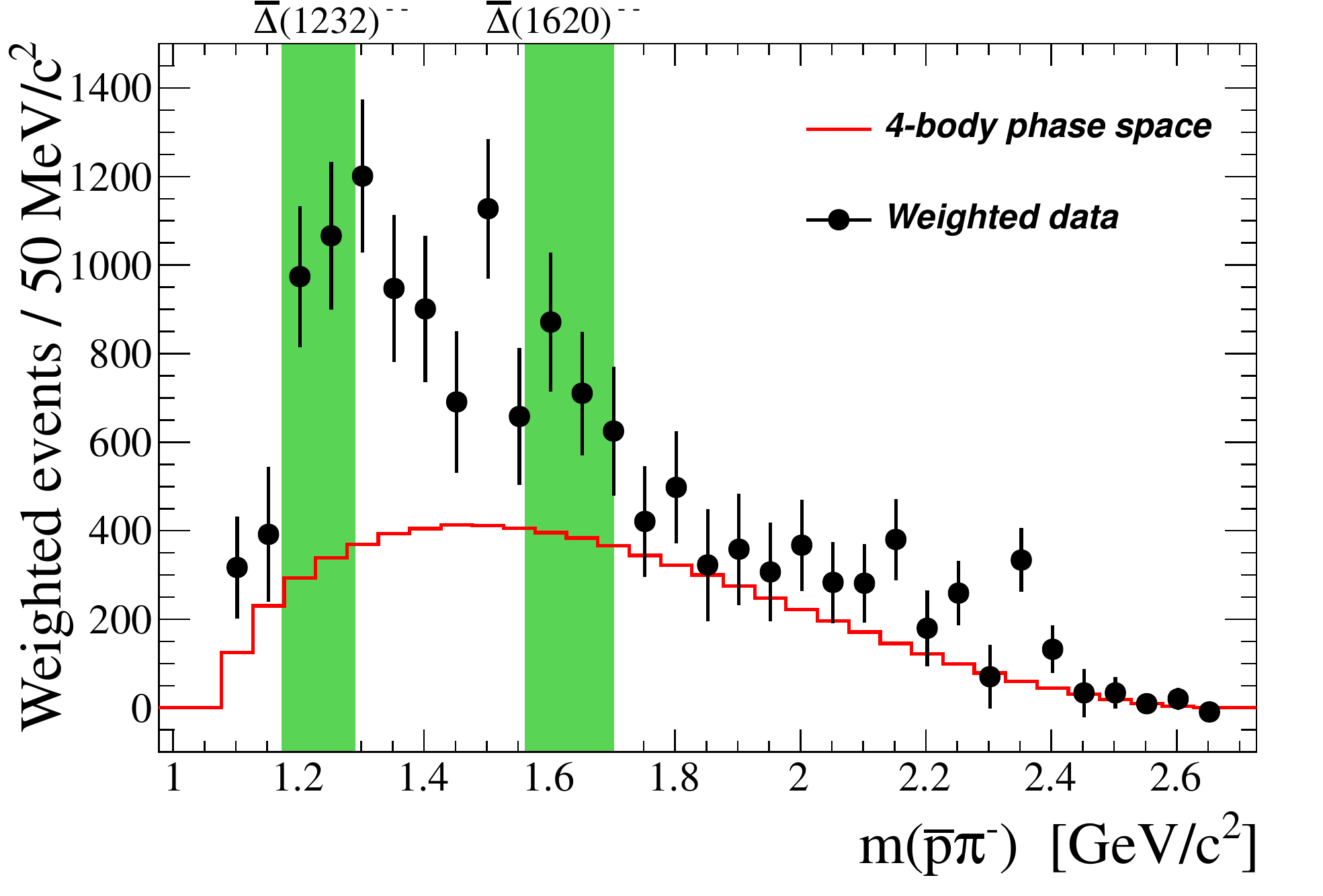}
	\caption{The $m(\antiproton\pim)$ distribution in data and simulated four-body phase space decays.
	The shaded vertical ranges represent a width of one $\Gamma$ and are centered 
	at the average mass of $\Deltabar^{--}(1232)$ and $\Deltabar^{--}(1620)$, respectively.
	The parameters are taken from the RPP \cite{ref:PDG}. The range of 
	$\Deltabar^{--}(1600)$ is not drawn since its parameters have large uncertainties.}
	\label{fig:5}
\end{figure}

\begin{figure}[htp]
	\centering\includegraphics[width=.49\textwidth]{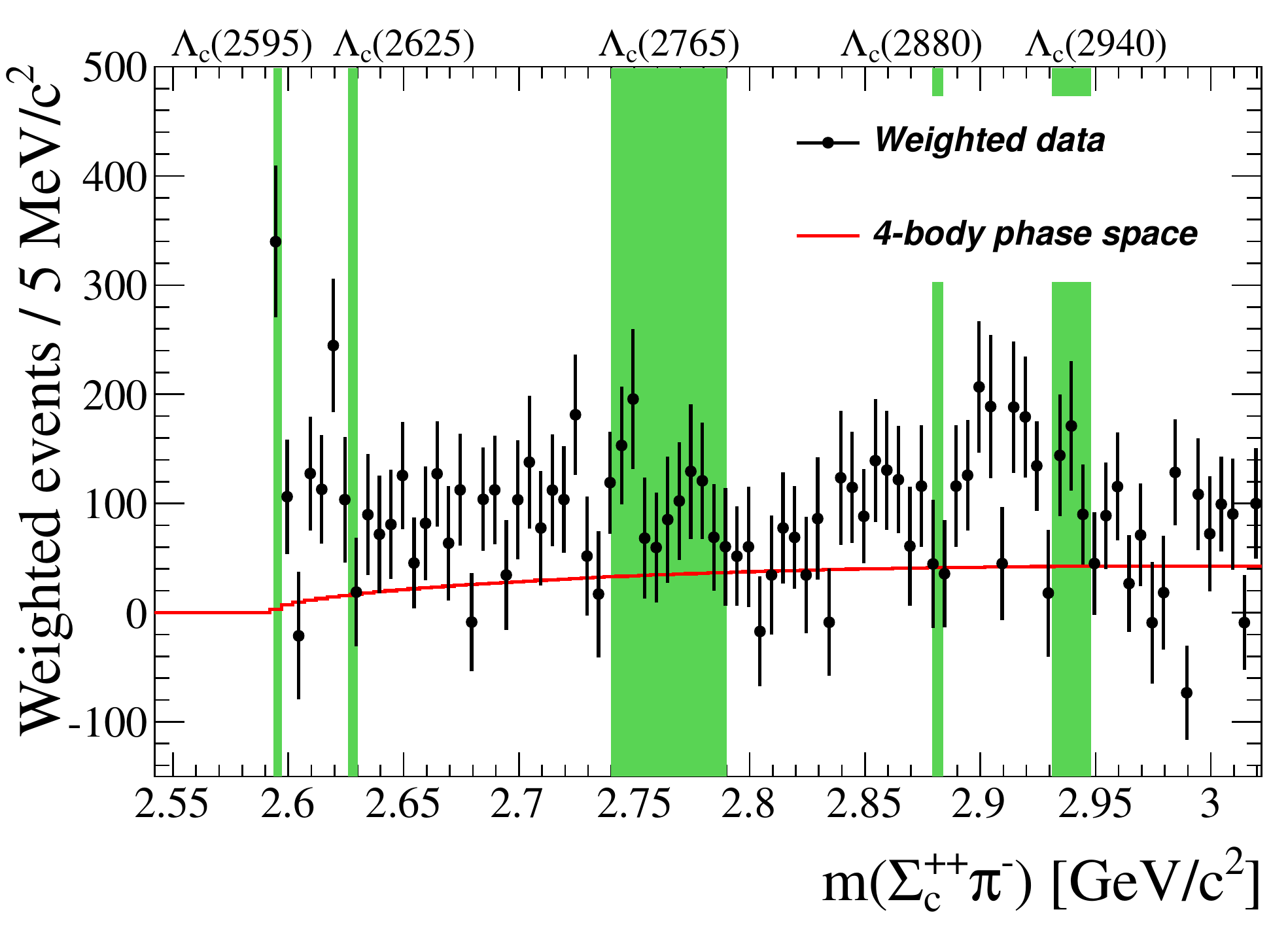}
	\caption{The $m(\Scpp\pim)$ distribution in data after efficiency correction and 
	\DeltaE-sideband subtraction. The solid line shows four-body phase space decays. The 
	shaded vertical ranges represent a width of one $\Gamma$ and are centered at the 
	average mass of the respective \Lcps resonance. The parameters are taken from the 
	RPP \cite{ref:PDG}.}
	\label{fig:6}
\end{figure}

\begin{figure}[htp]
	\centering\includegraphics[width=.5\textwidth]{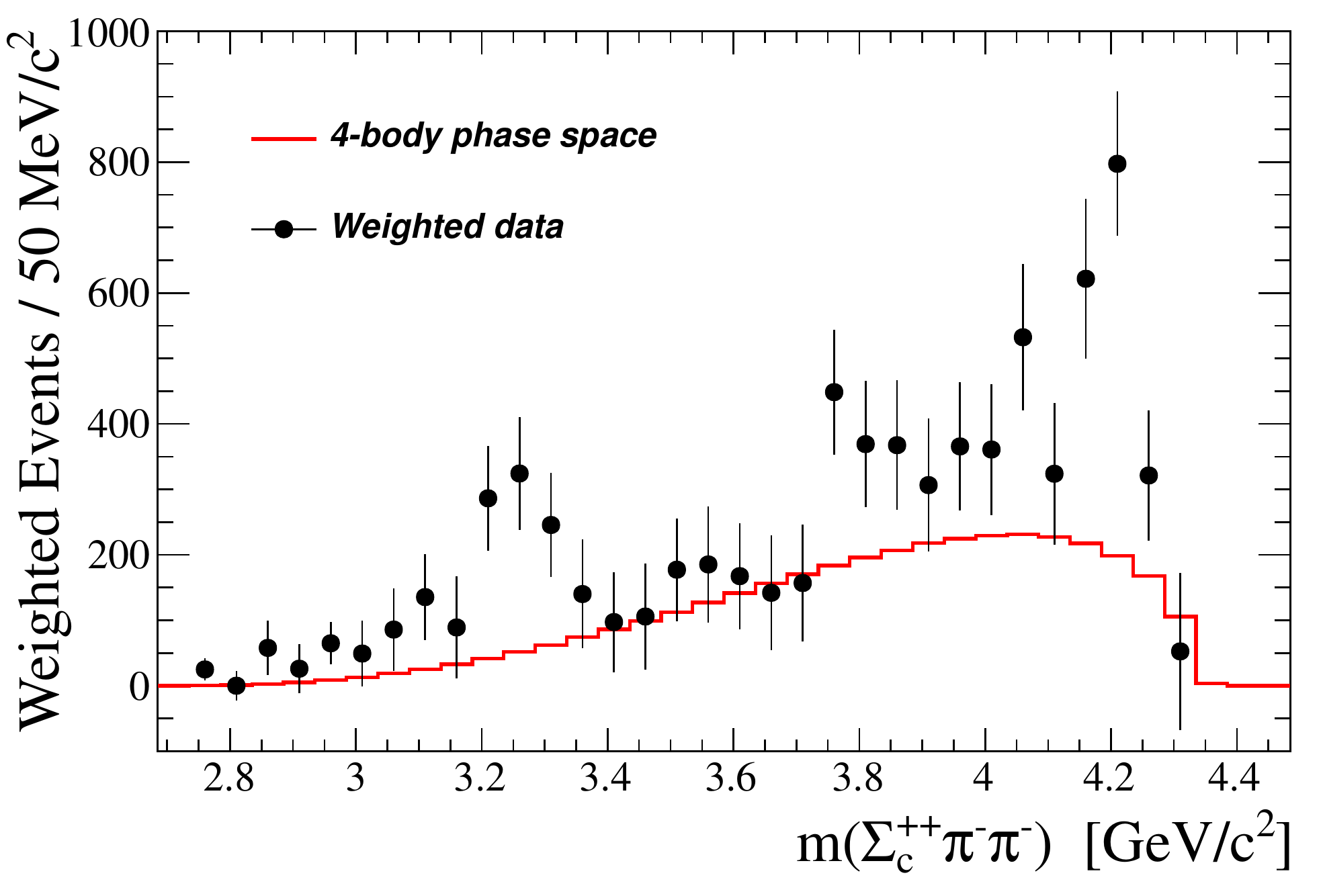}
	\caption{The $m(\Scpp\pim\pim)$ distribution in data and simulated four-body phase space decays.
	The histogram in data includes efficiency correction and \DeltaE-sideband subtraction 
	according to the definition in Fig. \ref{fig:2}.}
	\label{fig:7}
\end{figure}

\noindent
This percentage will be used to normalize the PS projection in the two-body and 
three-body invariant mass distributions in Figs. \ref{fig:5}\textendash\ref{fig:7}.


\vspace{1em}

\begin{center}
 \textbf{IX. Resonant subchannels}
\end{center}

Figure~\ref{fig:5} shows the invariant mass distribution of $\antiproton\pim=\{\antiproton\pislow,\antiproton\pifast\}$ 
[sum of the distributions of $m(\antiproton\pislow)$ and $m(\antiproton\pifast)$]
after sideband subtraction in \DeltaE (see Fig. \ref{fig:2} for the definition of the sidebands) 
and efficiency correction. The efficiency correction here and in the other invariant 
masses of the \B daughters is determined from PS MC for the particular mass that is considered.
The differences between data and PS in the range $m(\antiproton\pim)\in(1.2,1.7)\gevcc$ 
are compatible with the existence of the resonances $\Deltabar^{--}(1232,1600,1620)$.

\noindent
Figure~\ref{fig:6} shows the invariant mass of $\Scpp\pim=\{\Scpp\pislow,\Scpp\pifast\}$ after
efficiency correction and sideband subtraction in \DeltaE. The large number of events at
threshold are consistent with the decay $\Bub\to\Lambda_{c}(2595)^{+}\antiproton\pim$. 
There are no significant signals for other \Lcps resonances.

\noindent
In the three-body invariant mass distribution $m(\Scpp\pim\pim)$ (Fig. \ref{fig:7}) we see 
unexplained structures at $3.25\gevcc$, $3.8\gevcc$, and $4.2\gevcc$. However, because of the 
limited number of signal candidates, it is not possible to analyze these enhancements in more detail.

We find no indication of a threshold enhancement in the baryon-antibaryon mass distribution.

\begin{center}
 \textbf{X. Summary and Conclusions}
\end{center}

We have measured the branching fraction 
$\mybr=(2.98 \pm 0.22 \pm 0.77_{(\Lambda_c)}) \times 10^{-4}$.
This improves on the previous measurement by CLEO \cite{ref:CLEO}.

We have calculated an effective PS fraction of $49\,\%$ for the observed decay, which 
may indicate the importance of resonant substructures in baryonic \B decays. 
By comparing the data and four-body PS in the distributions of the invariant masses of 
the \B daughters, we find suggestions for the resonant subchannels
$\Bub\to\LcpsI\antiproton\pim$ and, possibly, $\mydecresA$. Additionally, we see 
unexplained structures in $m(\Scpp\pim\pim)$ at $3.25\gevcc$, $3.8\gevcc$, and $4.2\gevcc$.

Combining our measurement with the results \\
$\BR(\mydecS)=(4.4\pm2.0)\times10^{-4}$ and \\
$\BR(\mydecF)=(22.5\pm6.8)\times10^{-4}$ from CLEO \cite{ref:CLEO}, we calculate the 
resonant fractions
$\frac{\mybr}{\BR(\mydecF)}=(13.2\pm4.1)\,\%\,$ and \\
$\frac{\mybr + \BR(\mydecS)}{\BR(\mydecF)}=(33\pm13)\,\%\,$.

\vspace{1.0em}

\begin{center}
 \textbf{XI. Acknowledgments}
\end{center}


We are grateful for the 
extraordinary contributions of our \pep2 \ colleagues in
achieving the excellent luminosity and machine conditions
that have made this work possible.
The success of this project also relies critically on the 
expertise and dedication of the computing organizations that 
support \babar.
The collaborating institutions wish to thank 
SLAC for its support and the kind hospitality extended to them. 
This work is supported by the
US Department of Energy
and National Science Foundation, the
Natural Sciences and Engineering Research Council (Canada),
the Commissariat \`a l'Energie Atomique and
Institut National de Physique Nucl\'eaire et de Physique des Particules
(France), the
Bundesministerium f\"ur Bildung und Forschung and
Deutsche Forschungsgemeinschaft
(Germany), the
Istituto Nazionale di Fisica Nucleare (Italy),
the Foundation for Fundamental Research on Matter (The Netherlands),
the Research Council of Norway, the
Ministry of Education and Science of the Russian Federation, 
Ministerio de Ciencia e Innovaci\'on (Spain), and the
Science and Technology Facilities Council (United Kingdom).
Individuals have received support from 
the Marie-Curie IEF program (European Union) and the A. P. Sloan Foundation (USA).




\begin{thebibliography}{99}

\bibitem{ref:PDG}
K. Nakamura {\em et al.} (Particle Data Group), \jpg{37}, 075021 (2010).

\bibitem{footnote}
Throughout this paper, all decay   modes represent that mode and its charge conjugate.

\bibitem{ref:CLEO}
S.~A.~Dytman {\em et al.} (CLEO Collaboration),
  Phys.\ Rev.\ {\bf D66}, 091101 (2002).

\bibitem{ref:Belle2}
  K.~S.~Park {\it et al.}  (Belle Collaboration),
  Phys.\ Rev.\  D {\bf 75}, 011101 (2007).

\bibitem{ref:EvtGen}
  D.~J.~Lange,
  Nucl.\ Instrum.\ Meth.\ {\bf A462}, 152 (2001).

\bibitem{ref:geant}
S. Agostinelli {\em et al.} (GEANT4 Collaboration),
Nucl. Instrum. Methods {\bf A506}, 250 (2003).

\bibitem{ref:NIM}
B.\ Aubert {\em et al.} (\babar\ Collaboration), Nucl.\ Instrum.\ Methods {\bf A479}, 1 (2002).

\bibitem{Aubert:2002rg}
  B.~Aubert {\it et al.}  (BABAR Collaboration),
  Phys.\ Rev.\  D {\bf 66}, 032003 (2002).

\bibitem{babar:trackfinding}
  T.~Allmendinger {\it et al.},
  arXiv:1207.2849 [hep-ex] (submitted to Nucl. Instrum. Methods).

\end{thebibliography}
\end{document}